\documentclass[english,conference]{IEEEtran}
\usepackage[T1]{fontenc}
\usepackage[latin9]{inputenc}
\usepackage{verbatim}
\usepackage{units}
\usepackage{amsthm}
\usepackage{amsmath}
\usepackage{amssymb}
\usepackage{esint}

\makeatletter
\theoremstyle{definition}
\newtheorem{assumption}{Assumption}
\theoremstyle{definition}
\newtheorem{property}{Property}
  \theoremstyle{plain}
  \newtheorem{thm}{\protect\theoremname}
  \theoremstyle{remark}
  \newtheorem{rem}{\protect\remarkname}

\usepackage{cite}
\IEEEoverridecommandlockouts

\makeatother

\usepackage{babel}
\providecommand{\remarkname}{Remark}
\providecommand{\theoremname}{Theorem}

\begin{document}

\title{Online Approximate Optimal Station Keeping of an Autonomous Underwater
Vehicle%
\thanks{Patrick Walters and Warren E. Dixon are with
the Department of Mechanical and Aerospace Engineering, University
of Florida, Gainesville, FL, USA. Email: \{walters8, 
wdixon\}@ufl{}.edu%
}%
\thanks{This research is supported in part by NSF award numbers 0901491, 1161260,
1217908, ONR grant number N00014-13-1-0151, and a contract with the
AFRL Mathematical Modeling and Optimization Institute. Any opinions,
findings and conclusions or recommendations expressed in this material
are those of the authors and do not necessarily reflect the views
of the sponsoring agency.%
}}

\author{Patrick Walters and Warren E. Dixon}
\maketitle
\begin{abstract}
Online approximation of an optimal station keeping strategy for a
fully actuated six degrees-of-freedom autonomous underwater vehicle
is considered. The developed controller is an approximation of the
solution to a two player zero-sum game where the controller is the
minimizing player and an external disturbance is the maximizing player.
The solution is approximated using a reinforcement learning-based
actor-critic framework. The result guarantees uniformly ultimately
bounded (UUB) convergence of the states and UUB convergence of the
approximated policies to the optimal polices without the requirement
of persistence of excitation.
\end{abstract}

\section{Introduction}

Autonomous underwater vehicles (AUVs) play an increasingly important
role in commercial and military objectives. The operational tasks
of AUVs vary, including: inspection, monitoring, exploration, and
surveillance \cite{Griffiths2003}. During a mission, an AUV may be
required to remain on station for an extended period of time, e.g.,
as a communication link for multiple vehicles, or for persistent environmental
monitoring of a specific area. The success of the mission could rely
on the vehicle's ability to hold a precise station (e.g., station
keeping near underwater structures and features) while maximizing
its time on station. Energy expended for propulsion is tightly coupled
to the endurance of AUVs \cite{Bellingham2010}, specially when station
keeping in extreme environments with strong currents or high seas.
Therefore, by reducing the energy expended for extended station keeping,
the time on station can be maximized.

The precise station keeping of an AUV is challenging because of the
six degree-of-freedom (DOF) nonlinear dynamics of the vehicle and
unmodeled environmental disturbances, such as surface effects and
ocean currents. Common approaches to the control of an underwater
vehicle include robust and adaptive control methods \cite{Yoerger1985,Healey1993a,Lapierre2008,Sebastian2007}.
These methods provide robustness to disturbances or model uncertainty;
however, do not explicitly attempt to reduce energy expended by propulsion.
This motivates the use of control methods where an optimal control
policy can be selected to satisfy a performance criteria, such as
the infinite-horizon quadratic performance criteria used to develop
optimal policies that minimize the square of the total control effort
(energy expended) and state error (accuracy) \cite{Kirk2004}. Because
of the difficulties associated with finding closed-form analytical
solutions to optimal control problems for nonlinear systems, previous
results in literature have linearized the AUV model, developing $\mathrm{H_{\infty}}$
control policies \cite{Kaminer1991,Petrich2011} and a model-based
predictive control policy \cite{Riedel2005}. Considering the nonlinear
AUV dynamics, \cite{McLain1998} numerically approximates the solution
to the Hamilton-Jacobi-Bellman equation using an iterative application
of Galerkin's method and \cite{Park2000b} develops a PID-based $\mathrm{H}_{\infty}$
control strategy.

Reinforcement learning-based (RL-based) methods have been recently
used to approximate solutions to optimal control problems \cite{Vamvoudakis2010,Vamvoudakis2010a,Bhasin.Kamalapurkar.ea2013a,Johnson2011a}.
The online RL-based actor-critic framework in \cite{Bhasin.Kamalapurkar.ea2013a}
and \cite{Johnson2011a} approximate the value function and optimal
policies of an optimal control problem using the so-called Bellman
error. The actor-critic framework approximation provides uniformly
ultimately bounded (UUB) convergence of the state to the origin and
UUB convergence of the approximate policy to the optimal policy. These
methods require persistence of excitation (PE) to insure convergence
to the optimal policy, which is undesirable for an operational underwater
vehicle.

In this result, a two player zero-sum differential game is developed
where the controller is the minimizing player and an external disturbance
is the maximizing player. The performance criteria for the two player
game is from the $\mathrm{H_{\infty}}$ control problem. This performance
criteria captures the desire to develop an optimal policy similar
to the infinite-horizon quadratic performance criteria, yet still
including the need for attenuating the unmodeled disturbances. The
developed controller differs from results such as \cite{Vamvoudakis2010a}
and \cite{Bhasin.Kamalapurkar.ea2013a} in that it removes the PE
requirement through the addition of concurrent-learning to the value
function's adaptive update law. As outlined in \cite{Chowdhary2010},
concurrent-learning uses the additional knowledge of recorded data
to remove the PE requirement. Due to the unique structure of the actor-critic
framework, the recorded data is replaced with sampled data points
at the current time instant. This paper presents a novel approach
to station keeping of a fully actuated 6 DOF AUV, which is robust
to unmodeled environmental disturbances using the actor-critic framework
to approximate a solution to the two player zero-sum game without
the need for PE. A Lyapunov-based stability analysis is presented
which guarantees UUB convergence of the states and UUB convergence
of the approximated policies to the optimal policies.

\section{Vehicle Model}

Consider the nonlinear equations of motion for an underwater vehicle
with the addition of an unknown additive disturbance given by \cite{Fossen2011}
\begin{equation}
\dot{\eta}=J\left(\eta\right)\nu,\label{eq:kinematics}
\end{equation}
\begin{equation}
M\dot{\nu}+C\left(\nu\right)\nu+D\left(\nu\right)\nu+g\left(\eta\right)=\tau_{b}+\tau_{d},\label{eq:dynamics}
\end{equation}
where $\nu\in\mathbb{R}^{6}$ is the body-fixed translational and
angular velocity vector, $\eta\in\mathbb{R}^{6}$ is the earth-fixed
position and orientation vector, $J:\mathbb{R}^{6}\rightarrow\mathbb{R}^{6\times6}$
is the coordinate transformation between the body-fixed and earth-fixed
coordinates, $M\in\mathbb{R}^{6\times6}$ is the inertia matrix including
added mass, $C:\mathbb{R}^{6}\rightarrow\mathbb{R}^{6\times6}$ is
the centripetal and Coriolis matrix, $D:\mathbb{R}^{6}\rightarrow\mathbb{R}^{6\times6}$
is the hydrodynamic damping and friction matrix, $g:\mathbb{R}^{6}\rightarrow\mathbb{R}^{6}$
is the gravitational and buoyancy force and moment vector, $\tau_{d}\in\mathbb{R}^{6}$
is the unknown disturbance (e.g., ocean currents and surface effects),
and $\tau_{b}\in\mathbb{R}^{6}$ is the body-fixed force and moment
control input. The state vectors in $\left(\ref{eq:kinematics}\right)$
are further defined as
\[
\eta=\left[\begin{array}{cccccc}
x & y & z & \phi & \theta & \psi\end{array}\right]^{T},
\]
\[
\nu=\left[\begin{array}{cccccc}
u & v & w & p & q & r\end{array}\right]^{T},
\]
where $x$, $y$, $z\in\mathbb{R}$ are the earth-fixed position vector
components of the center of mass, $\phi$, $\theta$, $\psi\in\mathbb{R}$
represent the roll, pitch, and yaw angles, respectively, $u$, $v$,
$w\in\mathbb{R}$ are the body-fixed translational velocities, and
$p$, $q$, $r\in\mathbb{R}$ are the body-fixed angular velocities.
The coordinate transformation $J$ is defined as
\[
J\triangleq\left[\begin{array}{cc}
J_{1}\left(\eta\right) & 0_{3\times3}\\
0_{3\times3} & J_{2}\left(\eta\right)
\end{array}\right],
\]
where $J_{1}:\mathbb{R}^{6}\rightarrow\mathbb{R}^{3\times3}$ is the
coordinate transformation between body-fixed and earth-fixed translational
velocities represented by the Z-Y-X Euler angle rotation matrix as
\[
J_{1}\triangleq\left[\begin{array}{ccc}
c\psi c\theta & -s\psi c\phi+c\psi s\theta s\phi & s\psi s\phi+c\psi s\theta c\phi\\
s\psi c\theta & c\psi c\phi+s\psi s\theta s\phi & -c\psi s\phi+s\psi s\theta c\phi\\
-s\theta & c\theta s\phi & c\theta c\phi
\end{array}\right],
\]
and $J_{2}:\mathbb{R}^{6}\rightarrow\mathbb{R}^{3\times3}$ represents
the coordinate transformation between body-fixed and earth-fixed angular
velocities defined as 
\[
J_{2}\triangleq\left[\begin{array}{ccc}
1 & s\phi t\theta & c\phi t\theta\\
0 & c\phi & -s\phi\\
0 & s\phi/c\theta & c\phi/c\theta
\end{array}\right],
\]
where $s\cdot$, $c\cdot$, $t\cdot$ denote $\sin\left(\cdot\right)$,
$\cos\left(\cdot\right)$, $\tan\left(\cdot\right)$, respectively.
\begin{assumption}
\label{thm:pitch_const} A pitch of $\pm\frac{\pi}{2}\unit{rad}$
is avoided. 
\end{assumption}

For a vehicle with metacentric stability, Assumption \ref{thm:pitch_const}
is easily satisfied \cite{Fossen2011}; therefore, the coordinate
transformation $J$, and it's inverse exist and are bounded. By applying
the kinematic relationship in $\left(\ref{eq:kinematics}\right)$
to $\left(\ref{eq:dynamics}\right)$, the vehicle dynamics can be
expressed in the earth-fixed frame as \cite{Fossen2011}
\[
\bar{M}\left(\eta\right)\ddot{\eta}+\bar{C}\left(\eta,\dot{\eta},\nu\right)\dot{\eta}+\bar{D}\left(\eta,\nu\right)\dot{\eta}+\bar{g}\left(\eta\right)=\bar{\tau}_{b}+\bar{\tau}_{d},
\]
where $\bar{M}:\mathbb{R}^{6}\rightarrow\mathbb{R}^{6\times6}$, $\bar{C}:\mathbb{R}^{6}\times\mathbb{R}^{6}\times\mathbb{R}^{6}\rightarrow\mathbb{R}^{6\times6}$,
$\bar{D}:\mathbb{R}^{6}\times\mathbb{R}^{6}\rightarrow\mathbb{R}^{6\times6}$
and $\bar{g}:\mathbb{R}^{6}\rightarrow\mathbb{R}^{6}$, and are defined
as $\bar{M}\triangleq J^{-T}MJ^{-1},\:\bar{C}\triangleq J^{-T}\left[C-MJ^{-1}\dot{J}\right]J^{-1},\:\bar{D}\triangleq J^{-T}DJ^{-1},\:\bar{g}\triangleq J^{-T}g,\:\bar{\tau}_{b}\triangleq J^{-T}\tau_{b}$
and $\bar{\tau}_{d}\triangleq J^{-T}\tau_{d}$. 
\begin{property}
The transformed inertia matrix $\bar{M}$ is symmetric, positive definite
\cite{Fossen2011}, and satisfies $\underline{m}\left\Vert \xi_{m}\right\Vert ^{2}\leq\xi_{m}^{T}\bar{M}\left(\eta\right)\xi_{m}\leq\overline{m}\left(\eta\right)\left\Vert \xi_{m}\right\Vert ^{2},\:\forall\xi_{m},\eta\in R^{6}$
where $\underline{m}\in\mathbb{R}$ is a positive known constant,
and $\overline{m}:\mathbb{R}^{6}\rightarrow\left[0,\infty\right)$
is a positive known function. The inverse $\bar{M}^{-1}$ satisfies
$\frac{1}{\overline{m}\left(\eta\right)}\left\Vert \xi_{m}\right\Vert ^{2}\leq\xi_{m}^{T}\bar{M}^{-1}\left(\eta\right)\xi_{m}\leq\frac{1}{\underline{m}}\left\Vert \xi_{m}\right\Vert ^{2},\:\forall\xi_{m},\:\eta\in R^{6}$.
\end{property}

For the subsequent development, $\eta$ and $\nu$ are assumed to
be measurable by sensors commonly used for underwater navigation,
such as the ones found in \cite{Kinsey2006}. The matrices $J$, $\bar{M}$,
$\bar{C}$, $\bar{D}$, and $\bar{g}$ are assumed to be known, while
the additive disturbance $\bar{\tau}_{d}$ is assumed to be unknown.
The dynamics can be rewritten in the following control affine form:
\[
\dot{\zeta}=f\left(\zeta\right)+g\left(\zeta\right)\left(u_{1}+u_{2}\right),
\]
where $\zeta\triangleq\left[\begin{array}{cc}
\eta & \dot{\eta}\end{array}\right]^{T}\in\mathbb{R}^{12}$ is the state vector, $u_{1}\triangleq\bar{\tau}_{b}$, $u_{2}\triangleq\bar{\tau}_{d}\in\mathbb{R}^{6}$
are the control vectors, and the functions $f:\mathbb{R}^{12}\rightarrow\mathbb{R}^{12}$
and $g:\mathbb{R}^{12}\rightarrow\mathbb{R}^{12\times6}$ are locally
Lipschitz and defined as
\[
f\triangleq\left[\begin{array}{c}
\dot{\eta}\\
-\bar{M}^{-1}\bar{C}\dot{\eta}-\bar{M}^{-1}\bar{D}\dot{\eta}-\bar{M}^{-1}\bar{g}
\end{array}\right],
\]
\begin{equation}
g\triangleq\left[\begin{array}{c}
0\\
\bar{M}^{-1}
\end{array}\right].\label{eq:adp_dynamics}
\end{equation}

\begin{assumption}
\label{thm:neutral}The underwater vehicle is neutrally buoyant and
the center of gravity is located vertically below the center of buoyancy
on the $z$ axis; hence, $f\left(0\right)=0$ due to the form of $\bar{g}$
\cite{Fossen2011}. 
\end{assumption}

\section{Formulation of Two Player Zero-Sum Differential Game}

The performance index for the $\mathrm{H_{\infty}}$ control problem
is \cite{Lewis1986}
\begin{equation}
J_{c}\left(\zeta,u_{1},u_{2}\right)=\intop_{t}^{\infty}r\left(\zeta\left(\tau\right),u_{1}\left(\tau\right),u_{2}\left(\tau\right)\right)d\tau,\label{eq:perform_index}
\end{equation}
where $r:\mathbb{R}^{12}\rightarrow\left[0,\infty\right)$ is the
local cost defined as
\begin{equation}
r\left(\zeta,u_{1},u_{2}\right)\triangleq\zeta^{T}Q\zeta+u_{1}^{T}Ru_{1}-\gamma^{2}u_{2}^{T}u_{2}.\label{eq:local_cost}
\end{equation}
In $\left(\ref{eq:local_cost}\right)$, $Q\in\mathbb{R}^{12\times12}$
is positive definite matrix, $R\in\mathbb{R}^{6\times6}$ is symmetric
positive definite matrix, $u_{1}$ is the controller and the minimizing
player, $u_{2}$ is the disturbance and the maximizing player, $\gamma\geq\gamma^{*}>0$
where $\gamma^{*}$ is the smallest value of $\gamma$ for which the
system is stabilized \cite{van1992l2}. The matrix $Q$ has the property
$\underline{q}\left\Vert \xi_{q}\right\Vert ^{2}\leq\xi_{q}^{T}Q\xi_{q}\leq\overline{q}\left\Vert \xi_{q}\right\Vert ^{2},\:\forall\xi_{q}\in\mathbb{R}^{12}$
where $\underline{q}$ and $\overline{q}$ are positive constants.
The infinite-time scalar value functional $V:\left[0,\infty\right)\rightarrow\left[0,\infty\right)$
for the two player zero-sum game is written as
\[
V=\underset{u_{1}}{\min}\underset{u_{2}}{\max}\intop_{t}^{\infty}r\left(\zeta\left(\tau\right),u_{1}\left(\tau\right),u_{2}\left(\tau\right)\right)d\tau,
\]
A unique solution exists to the differential game if the Nash condition
holds
\[
\underset{u_{1}}{\min}\underset{u_{2}}{\max}J_{c}\left(\zeta\left(0\right),u_{1},u_{2}\right)=\underset{u_{1}}{\max}\underset{u_{2}}{\min}J_{c}\left(\zeta\left(0\right),u_{1},u_{2}\right).
\]

The objective of the optimal control problem is to find the optimal
policies $u_{1}^{*}$ and $u_{2}^{*}$ that minimize the performance
index $\left(\ref{eq:perform_index}\right)$ subject to the dynamic
constraints in $\left(\ref{eq:adp_dynamics}\right)$. Assuming that
a minimizing policy exists and the value function is continuously
differentiable, the Hamiltonian is defined as
\begin{equation}
H\triangleq r\left(\zeta,u_{1}^{*},u_{2}^{*}\right)+\frac{\partial V}{\partial\zeta}\left(f+g\left(u_{1}^{*}+u_{2}^{*}\right)\right),\label{eq:ham}
\end{equation}
The Hamilton-Jacobi-Isaac\textquoteright{}s (HJI) equation is given
as \cite{BasarGame}
\begin{equation}
0=\frac{\partial V}{\partial t}+H,\label{eq:HJI}
\end{equation}
where $\frac{\partial V}{\partial t}\equiv0$ since the value function
is not an explicit function of time. After substituting $\left(\ref{eq:local_cost}\right)$
into $\left(\ref{eq:HJI}\right)$ , the optimal policies are given
by \cite{Kirk2004}
\begin{equation}
u_{1}^{*}=-\frac{1}{2}R^{-1}g^{T}\left(\frac{\partial V}{\partial\zeta}\right)^{T},\label{eq:u1_opt}
\end{equation}
\begin{equation}
u_{2}^{*}=\frac{1}{2\gamma^{2}}g^{T}\left(\frac{\partial V}{\partial\zeta}\right)^{T}.\label{eq:u2_opt}
\end{equation}

The analytical expressions for the optimal controllers in $\left(\ref{eq:u1_opt}\right)$
and $\left(\ref{eq:u2_opt}\right)$ require knowledge of the value
function which is the solution to the HJI equation in $\left(\ref{eq:HJI}\right)$.
A closed-form analytical solution to the HJI equation is generally
infeasible; hence, the subsequent development seeks an approximate
solution.

\section{Approximate Solution}

While various function approximation methods could be used, the subsequent
development is based on the use of neural networks (NNs) to approximate
the value function and optimal policies. The subsequent development
is also based on a temporary assumption that the state lies on a compact
set where $\zeta\left(t\right)\in\chi\subset\mathbb{R}^{12},\:\forall t\in\left[0,\infty\right)$.
This assumption is common in NN literature (cf. \cite{Hornik1990,Lewis2002a}),
and is relieved by the subsequent stability analysis (Remark \ref{rmk: semi-global}).
Specifically, a semi-global analysis indicates that if the initial
state is bounded, then the entire state trajectory remains on a compact
set.
\begin{assumption}
\label{thm:value_NN}The value function can be represented by a single-layer
NN with $m$ neurons as
\begin{equation}
V=W^{T}\sigma+\epsilon,\label{eq:value_NN}
\end{equation}
where $W\in\mathbb{R}^{m}$ is the ideal weight vector bounded above
by a known positive constant, $\sigma:\mathbb{R}^{12}\rightarrow\mathbb{R}^{m}$
is a bounded, continuously differentiable activation function, and
$\epsilon:\mathbb{R}^{12}\rightarrow\mathbb{R}$ is the bounded, continuously
differential function reconstruction error. 
\end{assumption}

Using $\left(\ref{eq:u1_opt}\right)$-$\left(\ref{eq:value_NN}\right)$,
the optimal policies can be represented as
\begin{equation}
u_{1}^{*}=-\frac{1}{2}R^{-1}g^{T}\left(\sigma'{}^{T}W+\epsilon'{}^{T}\right),\label{eq:u1_NN}
\end{equation}
\begin{equation}
u_{2}^{*}=\frac{1}{2\gamma^{2}}g^{T}\left(\sigma'{}^{T}W+\epsilon'{}^{T}\right).\label{eq:u2_NN}
\end{equation}
Based on $\left(\ref{eq:value_NN}\right)$-$\left(\ref{eq:u2_NN}\right)$,
NN approximations of the value function and the optimal policy are
defined as
\begin{equation}
\hat{V}=\hat{W}_{c}^{T}\sigma,\label{eq:V_approx}
\end{equation}
\begin{equation}
\hat{u}_{1}=-\frac{1}{2}R^{-1}g^{T}\sigma'^{T}\hat{W}_{a1},\label{eq:u1_approx}
\end{equation}
\begin{equation}
\hat{u}_{2}=\frac{1}{2\gamma^{2}}g^{T}\sigma'^{T}\hat{W}_{a2},\label{eq:u2_approx}
\end{equation}
where $\hat{W}_{c},\hat{W}_{a1},\hat{W}_{a2}\in\mathbb{R}^{m}$ are
estimates of the constant ideal weight vector $W$. The weight estimation
errors are defined as $\tilde{W}_{c}\triangleq W-\hat{W}_{c}$, $\tilde{W}_{a1}\triangleq W-\hat{W}_{a1}$,
and $\tilde{W}_{a2}\triangleq W-\hat{W}_{a2}$. Substituting $\left(\ref{eq:V_approx}\right)$-$\left(\ref{eq:u2_approx}\right)$
into $\left(\ref{eq:ham}\right)$, the approximate Hamiltonian is
given by
\begin{equation}
\hat{H}=r\left(\zeta,\hat{u}_{1},\hat{u}_{2}\right)+\frac{\partial\hat{V}}{\partial\zeta}\left(f+g\left(\hat{u}_{1}+\hat{u}_{2}\right)\right).\label{eq:HJI_approx}
\end{equation}
The error between the optimal and approximate Hamiltonian is called
the Bellman error $\delta\in\mathbb{R}$, defined as
\begin{equation}
\delta\triangleq\hat{H}-H,\label{eq:bellman_err}
\end{equation}
where $H\equiv0.$ Therefore, the Bellman error can be written in
a measurable form as
\[
\delta=r\left(\zeta,\hat{u}_{1},\hat{u}_{2}\right)+\hat{W}_{c}^{T}\omega,
\]
where $\omega\triangleq\sigma'\left(f+g\left(\hat{u}_{1}+\hat{u}_{2}\right)\right)\in\mathbb{R}^{m}$. 
\begin{assumption}
\label{thm:concurrent}There exists a set of sampled data points $\left\{ \zeta_{j}\in\chi|j=1,2,\ldots,N\right\} $
such that $\forall t\in\left[0,\infty\right)$,
\begin{equation}
\mathrm{rank}\left(\sum_{j=1}^{N}\frac{\omega_{j}\omega_{j}^{T}}{p_{j}}\right)=L,\label{eq:rank_cond}
\end{equation}
where $p_{j}\triangleq\sqrt{1+\omega_{j}^{T}\omega_{j}}$ the normalization
constant and $\omega_{j}$ are evaluated at the specified data point,
$\zeta_{j}$. 
\end{assumption}

The rank condition in $\left(\ref{eq:rank_cond}\right)$ cannot be
guaranteed to hold a priori. However, heuristically, the condition
can be met by sampling redundant data, i.e., $N\gg L$. Based on Assumption
\ref{thm:concurrent}, it can be shown that $\sum_{j=1}^{N}\frac{\omega_{j}\omega_{j}^{T}}{p_{j}}>0$
such that
\[
\underline{c}\left\Vert \xi_{c}\right\Vert ^{2}\leq\xi_{c}^{T}\left(\sum_{j=1}^{n}\frac{\omega_{j}\omega_{j}^{T}}{p_{j}}\right)\xi_{c}\leq\overline{c}\left\Vert \xi_{c}\right\Vert ^{2},\:\forall\xi_{c}\in\mathbb{R}^{4}
\]
even in the absence of persistent excitation \cite{Chowdhary.Johnson2011a,Chowdhary.Yucelen.ea2012}.

The value function update law is based on concurrent learning gradient
descent of the Bellman error given by \cite{Chowdhary2010}
\begin{equation}
\dot{\hat{W}}_{c}=-\eta_{c}\frac{1}{p}\frac{\partial\delta}{\partial\hat{W}_{c}}\delta-\eta_{c}\sum_{j=1}^{n}\frac{1}{p_{j}}\frac{\partial\delta_{j}}{\partial\hat{W}_{c}}\delta_{j},\label{eq:wc_dot}
\end{equation}
where $\eta_{c}\in\mathbb{R}$ is a positive adaptation gain, $\frac{\partial\delta}{\partial\hat{W}_{c}}=\omega$
is the regressor matrix, $p\triangleq\sqrt{1+\omega^{T}\omega}$ is
a normalization constant.

The policy NN update laws are given by 
\begin{equation}
\dot{\hat{W}}_{a1}=\mbox{proj}\left\{ -\eta_{a1}\left(\hat{W}_{a1}-\hat{W}_{c}\right)\right\} ,\label{eq:wa1_dot}
\end{equation}
\begin{equation}
\dot{\hat{W}}_{a2}=\mbox{proj}\left\{ -\eta_{a2}\left(\hat{W}_{a2}-\hat{W}_{c}\right)\right\} ,\label{eq:wa2_dot}
\end{equation}
where $\eta_{a1}$, $\eta_{a2}\in\mathbb{R}$ are positive gains,
and $\mbox{proj}\left\{ \cdot\right\} $ is a smooth projection operator
used to bound the weight estimates \cite{Dixon2003}. Using Assumption
\ref{thm:value_NN} and properties of the projection operator, the
policy NN weight estimation errors can be bounded above by positive
constants.

\section{Stability Analysis}

An unmeasurable form of the Bellman error can be written using $\left(\ref{eq:ham}\right)$,
$\left(\ref{eq:HJI_approx}\right)$ and $\left(\ref{eq:bellman_err}\right)$,
as%
\begin{eqnarray}
\delta & = & -\tilde{W}_{c}^{T}\omega-\epsilon'f+\frac{1}{4}\epsilon'G_{1}\epsilon'^{T}-\frac{1}{4}\epsilon'G_{2}\epsilon'^{T}\nonumber \\
 &  & +\frac{1}{2}\epsilon'G_{1}\sigma'{}^{T}W-\frac{1}{2}\epsilon'G_{2}\sigma'{}^{T}W\label{eq:unmeas_HJB}\\
 &  & +\frac{1}{4}\tilde{W}_{a1}^{T}G_{\sigma1}\tilde{W}_{a1}-\frac{1}{4}\tilde{W}_{a2}^{T}G_{\sigma2}\tilde{W}_{a2},\nonumber 
\end{eqnarray}
where $G_{1}\triangleq gR^{-1}g^{T}\in\mathbb{R}^{12\times12}$, $G_{2}\triangleq g\gamma^{-2}g^{T}\in\mathbb{R}^{12\times12}$,
$G_{\sigma1}\triangleq\sigma'G_{1}\sigma'{}^{T}\in\mathbb{R}^{m\times m}$
and $G_{\sigma2}\triangleq\sigma'G_{2}\sigma'{}^{T}\in\mathbb{R}^{m\times m}$
are symmetric, positive semi-definite matrices. Similarly, the Bellman
error at the sampled data points can be written as
\begin{eqnarray}
\delta_{j} & = & -\tilde{W}_{c}^{T}\omega_{j}+\frac{1}{4}\tilde{W}_{a1}^{T}G_{\sigma1j}\tilde{W}_{a1}\label{eq:unmeas_HJB_j}\\
 &  & -\frac{1}{4}\tilde{W}_{a2}^{T}G_{\sigma2j}\tilde{W}_{a2}+E_{j},\nonumber 
\end{eqnarray}
where $E_{j}\triangleq\frac{1}{2}\epsilon_{j}'\left(G_{1}-G_{2}\right)\sigma_{j}'{}^{T}W+\frac{1}{4}\epsilon_{j}'\left(G_{1}-G_{2}\right)\epsilon_{j}'^{T}-\epsilon_{j}'f_{j}\in\mathbb{R}$
is a constant at each data point. 

For the subsequent analysis, the function $f$ on the compact set
$\chi$ is Lipschitz continuous and can be bounded by
\[
\left\Vert f\left(\zeta\right)\right\Vert \leq L_{f}\left\Vert \zeta\right\Vert ,\:\forall\zeta\in\chi,
\]
where $L_{f}$ is a positive constant, and the normalized regressor
in $\left(\ref{eq:wc_dot}\right)$ can be upper bounded by $\left\Vert \frac{\omega}{p}\right\Vert \leq1.$
\begin{thm}
If Assumptions \ref{thm:pitch_const}-\ref{thm:value_NN} hold and
the following sufficient conditions are satisfied
\begin{equation}
\underline{q}>\frac{\eta_{c}L_{f}\overline{\left\Vert \epsilon'\right\Vert }\varepsilon}{2},\label{eq:Q_sc}
\end{equation}
\textup{
\begin{equation}
\underline{c}>\frac{L_{f}\overline{\left\Vert \epsilon'\right\Vert }}{2\varepsilon}+\frac{\eta_{a1}+\eta_{a2}}{2\eta_{c}},\label{eq:Wc_sc}
\end{equation}
}
\begin{equation}
\lambda_{\min}\left(R\right)\geq\gamma^{2},\label{eq:r_gam_cond}
\end{equation}
where $\overline{\left\Vert \cdot\right\Vert }\triangleq\sup_{\zeta}\left\Vert \cdot\right\Vert $
and $\lambda_{\min}\left(\cdot\right)$ represents the minimum eigenvalue,
and $Z\triangleq\left[\begin{array}{cccc}
\zeta^{T} & \tilde{W}_{c}^{T} & \tilde{W}_{a1}^{T} & \tilde{W}_{a2}^{T}\end{array}\right]^{T}\in\mathbb{R}^{12+3m}$, then the policies in $\left(\ref{eq:u1_approx}\right)$ and $\left(\ref{eq:u2_approx}\right)$
with the NN update laws in $\left(\ref{eq:wc_dot}\right)$-$\left(\ref{eq:wa2_dot}\right)$
guarantee UUB regulation of the state $\zeta\left(t\right)$ and UUB
convergence of the approximated policies $\hat{u}_{1}$ and $\hat{u}_{2}$
to the optimal policies $u_{1}^{*}$ and $u_{2}^{*}$.\end{thm}
\begin{IEEEproof}
Consider the continuously differentiable, positive definite candidate
Lyapunov function
\[
V_{L}=V+\frac{1}{2}\tilde{W_{c}}^{T}\tilde{W}_{c}+\frac{1}{2}\tilde{W}_{a1}^{T}\tilde{W}_{a1}+\frac{1}{2}\tilde{W}_{a2}^{T}\tilde{W}_{a2},
\]
where $V$ is positive definite when the sufficient condition in $\left(\ref{eq:r_gam_cond}\right)$
is satisfied. Since $V$ is continuously differentiable and positive
definite, from Lemma 4.3 of \cite{Khalil2002} there exist two class
$\mathcal{K}$ functions, such that
\begin{equation}
\alpha_{1}\left(\left\Vert \zeta\right\Vert \right)\leq V\left(\zeta\right)\leq\alpha_{2}\left(\left\Vert \zeta\right\Vert \right).\label{eq:value_classK}
\end{equation}
Using $\left(\ref{eq:value_classK}\right)$, $V_{L}$ can be bounded
by
\begin{equation}
\alpha_{3}\left(\left\Vert Z\right\Vert \right)\leq V_{L}\left(Z\right)\leq\alpha_{4}\left(\left\Vert Z\right\Vert \right),\label{eq:VL_classK}
\end{equation}
where $\alpha_{3}$ and $\alpha_{4}$ are class $\mathcal{K}$ functions.
The time derivative of the candidate Lyapunov function is
\begin{equation}
\dot{V}_{L}=\frac{\partial V}{\partial\zeta}f+\frac{\partial V}{\partial\zeta}g\left(\hat{u}_{1}+\hat{u}_{2}\right)-\tilde{W}_{c}^{T}\dot{\hat{W}}_{c}-\tilde{W}_{a1}^{T}\dot{\hat{W}}_{a1}-\tilde{W}_{a2}^{T}\dot{\hat{W}}_{a2}.\label{eq:vl_dot_1}
\end{equation}
Using $\left(\ref{eq:HJI}\right)$, $\frac{\partial V}{\partial\zeta}f=-\frac{\partial V}{\partial\zeta}g\left(u_{1}^{*}+u_{2}^{*}\right)-r\left(\zeta,u_{1}^{*},u_{2}^{*}\right)$.
Then, 
\begin{eqnarray*}
\dot{V}_{L} & = & \frac{\partial V}{\partial\zeta}g\left(\hat{u}_{1}+\hat{u}_{2}\right)-\frac{\partial V}{\partial\zeta}g\left(u_{1}^{*}+u_{2}^{*}\right)-r\left(\zeta,u_{1}^{*},u_{2}^{*}\right)\\
 &  & -\tilde{W}_{c}^{T}\dot{\hat{W}}_{c}-\tilde{W}_{a1}^{T}\dot{\hat{W}}_{a1}-\tilde{W}_{a2}^{T}\dot{\hat{W}}_{a2}.
\end{eqnarray*}
Substituting $\left(\ref{eq:wc_dot}\right)$-$\left(\ref{eq:wa2_dot}\right)$
for $\dot{\hat{W}}_{c}$, $\dot{\hat{W}}_{a1}$, and $\dot{\hat{W}}_{a2}$,
respectively, yields
\begin{eqnarray*}
\dot{V}_{L} & = & -\zeta^{T}Q\zeta-u_{1}^{*T}Ru_{1}^{*}+\gamma^{2}u_{2}^{*T}u_{2}^{*}+\frac{\partial V}{\partial\zeta}g\left(\hat{u}_{1}+\hat{u}_{2}\right)\\
 &  & -\frac{\partial V}{\partial\zeta}g\left(u_{1}^{*}+u_{2}^{*}\right)+\tilde{W}_{c}^{T}\left[\eta_{c}\frac{\omega^{T}}{p}\delta+\eta_{c}\sum_{j=1}^{n}\frac{\omega_{j}^{T}}{p_{j}}\delta_{j}\right]\\
 &  & +\tilde{W}_{a1}^{T}\eta_{a1}\left(\hat{W}_{a1}-\hat{W}_{c}\right)+\tilde{W}_{a2}^{T}\eta_{a2}\left(\hat{W}_{a2}-\hat{W}_{c}\right).
\end{eqnarray*}
Using Young's inequality, $\left(\ref{eq:value_NN}\right)$-$\left(\ref{eq:u2_NN}\right)$,
$\left(\ref{eq:u1_approx}\right)$, $\left(\ref{eq:u2_approx}\right)$,
$\left(\ref{eq:unmeas_HJB}\right)$, $\left(\ref{eq:unmeas_HJB_j}\right)$,
and $\left(\ref{eq:r_gam_cond}\right)$ the Lyapunov derivative can
be upper bounded as%
\begin{eqnarray*}
\dot{V}_{L} & \leq & -\varphi_{\zeta}\left\Vert \zeta\right\Vert ^{2}-\varphi_{c}\left\Vert \tilde{W}_{c}\right\Vert ^{2}-\varphi_{a1}\left\Vert \tilde{W}_{a1}\right\Vert ^{2}\\
 &  & -\varphi_{a2}\left\Vert \tilde{W}_{a2}\right\Vert ^{2}+\kappa_{a1}\left\Vert \tilde{W}_{a1}\right\Vert +\kappa_{a2}\left\Vert \tilde{W}_{a2}\right\Vert \\
 &  & +\kappa_{c}\left\Vert \tilde{W}_{c}\right\Vert +\kappa,
\end{eqnarray*}
where
\[
\varphi_{\zeta}=\underline{q}-\frac{\eta_{c}L_{f}\overline{\left\Vert \epsilon'\right\Vert }\varepsilon}{2},
\]
\[
\varphi_{c}=\eta_{c}\left(\underline{c}-\frac{L_{f}\overline{\left\Vert \epsilon'\right\Vert }}{2\varepsilon}-\frac{\eta_{a1}+\eta_{a2}}{2\eta_{c}}\right),
\]
\[
\varphi_{a1}=\frac{\eta_{a1}}{2},
\]
\[
\varphi_{a2}=\frac{\eta_{a2}}{2},
\]
\begin{eqnarray*}
\kappa_{c} & = & \sup_{\zeta\in\chi}\left\Vert \frac{\eta_{c}}{4}\tilde{W}_{a1}^{T}G_{\sigma1}\tilde{W}_{a1}+\frac{\eta_{c}}{4}\sum_{j=1}^{n}\tilde{W}_{a1}^{T}G_{\sigma1j}\tilde{W}_{a1}\right.\\
 &  & -\frac{\eta_{c}}{4}\tilde{W}_{a2}^{T}G_{\sigma2}\tilde{W}_{a2}-\frac{\eta_{c}}{4}\sum_{j=1}^{n}\tilde{W}_{a2}^{T}G_{\sigma2j}\tilde{W}_{a2}\\
 &  & +\frac{\eta_{c}}{2}\epsilon'\left(G_{1}-G_{2}\right)\sigma'{}^{T}W+\frac{\eta_{c}}{4}\epsilon'\left(G_{1}-G_{2}\right)\epsilon'^{T}\\
 &  & \left.+\eta_{c}\sum_{j=1}^{n}E_{j}\right\Vert ,
\end{eqnarray*}
\begin{eqnarray*}
\kappa_{a1} & = & \sup_{\zeta\in\chi}\left\Vert \frac{1}{2}W^{T}G_{\sigma1}+\frac{1}{2}\epsilon'G_{1}\sigma'^{T}\right\Vert ,
\end{eqnarray*}

\begin{eqnarray*}
\kappa_{a2} & = & \sup_{\zeta\in\chi}\left\Vert -\frac{1}{2}W^{T}G_{\sigma2}-\frac{1}{2}\epsilon'G_{2}\sigma'^{T}\right\Vert ,
\end{eqnarray*}

\[
\kappa=\sup_{\zeta\in\chi}\left\Vert \frac{1}{4}\epsilon'G\epsilon'^{T}\right\Vert .
\]
The constants $\varphi_{\zeta},\varphi_{c},\varphi_{a1},$ and $\varphi_{a2}$
are positive if the inequalities
\[
\underline{q}>\frac{\eta_{c}L_{f}\overline{\left\Vert \epsilon'\right\Vert }\varepsilon}{2},
\]
\begin{equation}
\underline{c}>\frac{L_{f}\overline{\left\Vert \epsilon'\right\Vert }}{2\varepsilon}+\frac{\eta_{a1}+\eta_{a2}}{2\eta_{c}}\label{eq:sconds}
\end{equation}
are satisfied. Completing the squares, the upper bound on the Lyapunov
derivative can be written as
\begin{eqnarray*}
\dot{V}_{L} & \leq & -\varphi_{\zeta}\left\Vert \zeta\right\Vert ^{2}-\frac{\varphi_{c}}{2}\left\Vert \tilde{W}_{c}\right\Vert ^{2}-\frac{\varphi_{a1}}{2}\left\Vert \tilde{W}_{a1}\right\Vert ^{2}\\
 &  & -\frac{\varphi_{a2}}{2}\left\Vert \tilde{W}_{a2}\right\Vert ^{2}+\frac{\kappa_{c}^{2}}{2\varphi_{c}}+\frac{\kappa_{a1}^{2}}{2\varphi_{a1}}+\frac{\kappa_{a2}^{2}}{2\varphi_{a2}}\\
 &  & +\kappa,
\end{eqnarray*}
which can be further upper bounded as
\begin{equation}
\dot{V}_{L}\leq-\alpha_{5}\left\Vert Z\right\Vert ,\:\forall\left\Vert Z\right\Vert \geq K>0,\label{eq:uub}
\end{equation}
where
\[
K\triangleq\sqrt{\frac{\kappa_{c}^{2}}{2\alpha_{5}\varphi_{c}}+\frac{\kappa_{a1}^{2}}{2\alpha_{5}\varphi_{a1}}+\frac{\kappa_{a2}^{2}}{2\alpha_{5}\varphi_{a2}}+\frac{\kappa}{\alpha_{5}}}
\]
and $\alpha_{5}$ is a positive constant. Invoking Theorem 4.18 in
\cite{Khalil2002}, $Z$ is UUB. Based on the definition of $Z$ and
the inequalities in $\left(\ref{eq:VL_classK}\right)$ and $\left(\ref{eq:uub}\right)$,
$\zeta,\tilde{W}_{c},\tilde{W}_{a1},\tilde{W}_{a2}\in\mathcal{L}_{\infty}$.
From the definition of $W$ and the NN weight estimation errors, $\hat{W}_{c},\hat{W}_{a1},\hat{W}_{a2}\in\mathcal{L}_{\infty}$.
Using the policy update laws, $\dot{\hat{W}}_{a1},\dot{\hat{W}}_{a2}\in\mathcal{L}_{\infty}$.
It follows that $\hat{V},\hat{u}_{1},\hat{u}_{2}\in\mathcal{L}_{\infty}$.
From the dynamics in $\left(\ref{eq:adp_dynamics}\right)$, $\dot{\zeta}\in\mathcal{L}_{\infty}$.
By the definition in $\left(\ref{eq:bellman_err}\right)$, $\delta\in\mathcal{L}_{\infty}$.
By the definition of the normalized value function update law, $\dot{\hat{W}}_{c}\in\mathcal{L}_{\infty}$.
\end{IEEEproof}

\begin{rem}
\label{rmk: semi-global}If $\left\Vert Z\left(0\right)\right\Vert \geq K$,
then $\dot{V}_{L}\left(Z\left(0\right)\right)<0$. There exists an
$\varepsilon_{1}\in\left[0,\infty\right)$ such that $V_{L}\left(Z\left(\varepsilon_{1}\right)\right)<V_{L}\left(Z\left(0\right)\right)$.
Using $\left(\ref{eq:VL_classK}\right)$, $\alpha_{3}\left(\left\Vert Z\left(\varepsilon_{1}\right)\right\Vert \right)\leq V_{L}\left(\varepsilon_{1}\right)<\alpha_{4}\left(\left\Vert Z\left(0\right)\right\Vert \right)$.
Rearranging terms, $\left\Vert Z\left(\varepsilon_{1}\right)\right\Vert <\alpha_{3}^{-1}\left(\alpha_{4}\left(\left\Vert Z\left(0\right)\right\Vert \right)\right)$.
Hence, $Z\left(\varepsilon_{1}\right)\in\mathcal{L}_{\infty}$. It
can be shown by induction that $Z\left(t\right)\in\mathcal{L}_{\infty},\:\forall t\in\left[0,\infty\right)$
when $\left\Vert Z\left(0\right)\right\Vert \geq K$. Using a similar
argument when $\left\Vert Z\left(0\right)\right\Vert <K$, $\left\Vert Z\left(t\right)\right\Vert <\alpha_{3}^{-1}\left(\alpha_{4}\left(K\right)\right)$.
Therefore, $Z\left(t\right)\in\mathcal{L}_{\infty},\:\forall t\in\left[0,\infty\right)$
when $\left\Vert Z\left(0\right)\right\Vert <K$. Since $Z\left(t\right)\in\mathcal{L}_{\infty},\:\forall t\in\left[0,\infty\right)$,
the state, $\zeta$, is shown to lie on the compact set, $\chi$,
where $\chi\triangleq\left\{ \zeta\in\mathbb{R}^{12}|\left\Vert \zeta\right\Vert \leq\alpha_{3}^{-1}\left(\alpha_{4}\left(\max\left(\left\Vert Z\left(0\right)\right\Vert ,K\right)\right)\right)\right\} $.
\end{rem}

\section{Conclusion}

An online approximation of a robust optimal control strategy is developed
to enable station keeping by an AUV. Using the RL-based actor-critic
framework, the solution to the HJI equation is approximated. A gradient
descent adaptive update law with concurrent-learning approximates
the value function. A Lyapunov-based stability analysis concludes
UUB convergence of the states and UUB convergence of the approximated
policies to the optimal polices without the requirement of PE.

Future work includes numerical simulation of the developed controller,
and comparison to a numerical offline optimal solution to evaluated
performance. Since a model for a arbitrary AUV is often difficult
to determine, extending the result to include a state estimator could
relax the requirement of exact model knowledge. Also, the extension
of the developed technique to underactuated AUVs would open the result
to a much broader class of vehicles.

\bibliographystyle{IEEEtran}
\bibliography{encr,master,ncr}

\end{document}